\documentclass[a4paper,pdflatex]{article}

\usepackage{graphicx,amssymb}

\def\nn{\nonumber\\}

\begin{document}
\begin{center}

{\bf \Large A Complete Version of the Glauber Theory for Elementary
Atom -- Target Atom Scattering and Its Approximations}

\bigskip

{\bf Olga Voskresenskaya\footnote{E-mail: voskr@jinr.ru} and
Alexander Tarasov}

\smallskip

{\it \small Joint Institute for Nuclear Research, 141980 Dubna,
Moscow Region, Russia}

\end{center}

\smallskip

\begin{abstract}
{\noindent A general formalism of the Glauber theory for elementary
atom (EA) -- target atom (TA) scattering is developed. A
second-order approximation of its complete version is considered in
the framework of the optical-model perturbative approach. A
`potential' approximation of a second-order optical model is
formulated neglecting the excitation effects of the TA. Its accuracy
is evaluated within the second-order approximation for the complete
version of the Glauber EA--TA scattering theory.}
\end{abstract}

\section{\large Introduction}

The experiment DIRAC (DImeson Relativistic Atom Complex), now
underway at the Proton Synchrotron, CERN
\cite{Dirac1,Dirac2,Dirac3}, aims to observe relativistic
hydrogenlike EA \cite{EA}\footnote{Elementary atoms $A_{ab}$ are the
Coulomb bound states of two elementary particles $a$ and $b$, which
can be, e.g., hadrons.} consisting of $\pi^\pm$ and/or
$\pi^\mp$/$K^\mp$ mesons (dimesoatoms/hadronic atoms) in 24 GeV
proton-nucleus interactions and to measure with a high precision
their lifetime. The interaction of the relativistic dimesoatoms
(DMA) with the ordinary target atoms is of particular importance for
the experiment because the DMA--TA interaction cross sections
accuracy plays a significant part in extracting the dimesoatoms
lifetime. For the DIRAC experiment to be successful, the excitation
and ionization cross sections of the pionium ($A_{2\pi}$) should be
known with accuracy $1\%$ or better. It has been pointed that by
using only the Glauber cross sections, one will be able to reach the
desired $1\%$ level accuracy for the target atom charge of $Z>60$.

The applications of the Glauber theory had originally been confined
within high-energy nuclear physics and fundamental particle physics
\cite{Glauber1,Glauber2}. For the relatively low energies, the
Glauber model for the elastic nucleon scattering
 has been modified to take the Coulomb field effect into account
\cite{CMG}. In \cite{Thomas}  one can find a review of using a
conventional Glauber approximation in the `atomic collisions', i.e.
in the intermediate- and high-energy target-inelastic scattering of
structureless charged particles by neutral atoms (H, He and alkali
metal target atoms) (see also \cite{Thom}). The only paper
reflecting  the investigations on the matter was devoted to the
atom--atom collisions \cite{Byronetal}. The authors of
\cite{Byronetal} tried to derive an expression of the cross section
for H(2s) quenching in the H(2s)--He(1s$^2$) interaction within the
eikonal approximation using an effective potential. Nevertheless, no
general formalism has been developed in the work though.

In a number of papers \cite{Tarasov,Heim}, an eikonal approach is
developed  for the computation of the total excitation cross
sections $\sigma^{tot}_{coh}(i)= \sum_{f}\sigma_{i\to f}$ of the
relativistic hadronic atoms ($A^{}_{2\pi}$, $A^{}_{\pi
\scriptscriptstyle K}$, $A^{}_{\scriptscriptstyle K K}$) interacting
with a screened Coulomb potential of the ordinary target atom (Ti,
Ni, Pt, etc.). These eikonal DMA excitation cross sections for the
Coulomb DMA--TA interaction take into account all multiphoton
DMA--TA exchange processes. However, within this approximation all
possible TA excitations  in intermediate and/or final states are
completely neglected. In other words, this description is
essentially grounded on the assumption that the TA Coulomb potential
does not vary in the course of the DMA--TA interaction.
Consequently, the calculated cross sections of the coherent
interaction $\sigma^{tot}_{coh}$ were identified with the total
cross sections
$\sigma^{tot}=\sigma^{tot}_{coh}+\sigma^{tot}_{incoh}$, where
$\sigma^{tot}_{incoh}\approx 0$, within this approximation.

In the context of the DIRAC experiment, the incoherent part
$\sigma^{tot}_{incoh}$ of the total cross sections corresponds to
the scattering with excitations of the TA electrons from a ground
state to all possible exited states. It should be noted that the TA
nuclear excitations are not considered in frames of this paper,
because a lot more excitation energy is required exceeding the
energy range relevant to the dimesoatom--atom scattering
\cite{Basel}. Estimation of the ratio
$\sigma^{tot}_{incoh}/\sigma^{tot}_{coh}$ for the EA--TA scattering
was performed by authors of \cite{Basel,Pak} using a `no correlation
limit' in the first-order Born approximation. It is shown that while
the incoherent scattering contribution to the $A_{2\pi}$--TA
interaction is negligible \cite{Basel},  it can not be neglected in
the calculation of the total $A_{2e}$--TA interaction cross sections
$\sigma^{tot}(i,I)=\sum_{f}\sum_{F}\sigma_{i+I \to f+F}$ \cite{Pak}.
A detailed study of the target electrons influence on the $A_{2\pi}$
scattering through screening and incoherent effects is performed in
\cite{Basel} using the one-photon approximation. Some simplest
results concerning the role of the multi-photon exchanges in the
incoherent EA--TA interaction are reported in \cite{Proc}.

In this work, the eikonal approximation for the DMA target-elastic
scattering neglecting all possible TA excitations is extended to
reflect these effects  within a second-order optical model of the
Glauber theory for the EA--TA scattering. In Section 2  we develop a
general formalism of the Glauber theory \cite{Glauber1,Glauber2} for
the EA--TA interactions.  Section 3 considers a second-order
perturbation approximation of its full version, a relationship
between the developed formalism and the results obtained in
\cite{Tarasov,Heim} is established, too. In Section 4 we formulate a
`potential' approximation for the second-order optical model and
evaluate its accuracy.  The results of our analysis are considered
in the context of the DIRAC experiment. In conclusion we briefly sum
up our findings.

This work is devoted to the memory of my friend and co-author, a
remarkable human being and scientist Alexander Tarasov, who untimely
passed away on March 19th, 2011.

\section{\large Complete version of the Glauber theory for  EA--TA scattering}

The  amplitude of the EA--TA interactions can be represented as
 \begin{eqnarray}\label{gl1} A_{i+I \to f+F}(\mathbf{q})=\frac{i}{2\pi} \int d^2b\,
 \exp(i\mathbf{q}
 \mathbf{b})\,\, \Gamma_{i+I\to f+F}(\mathbf{b}), \end{eqnarray}
where  $\mathbf{q}=\mathbf{k}-\mathbf{k^{\prime}}$ is a
two-dimensional momentum transfer, $\mathbf{k}$ and
$\mathbf{k^{\prime}}$ are the initial and final momenta of the
incident EA. The integration is carried out over a plane
perpendicular to the incident direction; $\mathbf{b}$ is an
impact-parameter vector in this plane;
 $\Gamma_{i+I \to f+F}(\mathbf{b})$ is the so-called profile function.

We can get a general formulation of the problem by considering the
EA scattering on a system of $Z$ constituents with the coordinates
$\mathbf{r_1},\,\mathbf{r_2},\ldots\,,\mathbf{r_{\scriptscriptstyle
Z}}$\footnote{For the energy range relevant  to the dimesoatom--atom
scattering, $\mathbf{r_k}$ ($k=\overline{1,Z}$) is a position vector
of a TA electron.} and the projections on the plane of the impact
parameter
$\mathbf{s_1},\,\mathbf{s_2},\ldots\,,\mathbf{s_{\scriptscriptstyle
Z}}$. If we introduce the configuration spaces for the EA wave
functions $\psi_i(\mathbf{r})$, $\psi_f(\mathbf{r})$ and the wave
functions $\Psi_I(\{\mathbf{r_k}\})$, $\Psi_F(\{\mathbf{r_k}\})$ of
the TA constituents in the initial $i,I$ and the final $f,F$ states,
the profile function can be written as
\begin{eqnarray}\label{Gamma}\Gamma_{i+I \to f+F}(\mathbf{b})&=&\int d^3r
\,\psi_f^{\ast}(\mathbf{r})\psi_i(\mathbf{r})\int
\prod_{k=1}^{Z}d^3r_k
\,\Psi_F^{\ast}(\{\mathbf{r_k}\})\Psi_I(\{\mathbf{r_k}\})\end{eqnarray}
$$\times\left(1-S(\mathbf{b},\mathbf{s},\{\mathbf{s_k}\}\right)$$
with an interaction operator
\begin{eqnarray} 1-S(\mathbf{b},\mathbf{s},\{\mathbf{s_k}\})=1-
\exp[i\Phi(\mathbf{b},\mathbf{ s},\{\mathbf{s_k}\})] \end{eqnarray}
and a phase-shift function
\begin{eqnarray} \label{Phi}\Phi(\mathbf{b},\mathbf{s},\{\mathbf{s_k}\})=
Z\Delta\chi(\mathbf{b},\mathbf{s})-
\sum\limits_{k=1}^{Z}\Delta\chi(\mathbf{b}-\mathbf{s_k},\mathbf{s}),
\end{eqnarray}
where the EA constituents phase-shift difference
$\Delta\chi(\mathbf{b},\mathbf{s})$ can be represented as follows:
\begin{eqnarray}\Delta\chi(\mathbf{b},\mathbf{s})= \frac{\alpha}{\beta}
\int\limits_{-\infty}^{\infty}dz
 \left[\big\vert\mathbf{
R}+\mathbf{r}/2\big\vert^{-1} -\big\vert\mathbf{R}-\mathbf{
r}/2\big\vert^{-1} \right],
 \end{eqnarray}
\begin{eqnarray} \mathbf{R}=(\mathbf{b},z),\quad \mathbf{r}=(\mathbf{s},z),
\quad \mathbf{r_k}=(\mathbf{s_k},z_k). \end{eqnarray}
Here, $Z$ denotes the TA nuclear charge, $\alpha$ is a fine
structure constant, $\beta=v/c=1$, $v$ is the EA velocity in the
laboratory frame, $z$ is a direction of incidence, $\mathbf{R}$ is a
radius-vector from the center mass of the target atom to the EA
center mass, $\mathbf{r}$ is a radius-vector from one EA constituent
to another.

The amplitude (\ref{gl1}) is normalized by the relations
\begin{eqnarray} 4\pi\mathrm{Im} A_{i+I \to i+I}(0)=\sigma^{tot}(i,I), \;\vert
A_{i+I \to f+F}(\mathbf{q})\vert^2=d\sigma_{i+I \to
f+F}/dq_{\bot},\end{eqnarray}
where
\begin{eqnarray}\label{sum1} \sigma^{tot}(i,I)=\sigma^{tot}_{coh}(i,I)+
\sigma^{tot}_{incoh}(i,I) =\sum_{f}\sum_{F}\sigma_{i+I \to f+F},
\end{eqnarray}
\begin{eqnarray}\label{sum2} \sigma^{tot}_{coh}(i,I)= \sum_{f}\sigma_{i+I \to
f+I},\quad \sigma^{tot}_{incoh}(i,I)=\sum_{f}\sum_{F\ne
I}\sigma_{i+I \to f+F}, \end{eqnarray}
\begin{eqnarray}\sigma_{i+I \to f+F}=\int d^2q \;d\sigma_{i+I \to
f+F}/dq_{\bot}.\end{eqnarray}

To find the total cross sections for all types of collisions in
which EA and TA begin in the states $i$ and $I$, one should sum the
partial cross sections in  (\ref{sum1}) and (\ref{sum2})  over all
states $f$ and $F$. The summation is easily performed using the
completeness relations:
\begin{eqnarray}\sum_f
\psi_f(\mathbf{r})\psi_f^{\ast}(\mathbf{r'})=
\delta(\mathbf{r}-\mathbf{r'}),
\end{eqnarray}
\begin{eqnarray}
\sum_F\Psi_F(\{\mathbf{
r_k}\})\Psi_F^{\ast}(\{\mathbf{r_k}\})=\prod_{k=1}^{Z}
\delta(\mathbf{r_k}-\mathbf{r_k'}). \end{eqnarray}
Taking into account the expression
\begin{eqnarray} \sum_{f,F}\frac{1}{2\pi}\int d^2q^{}_1A_{i_1+I_1 \to
f+F}(\mathbf{q^{}_1})A^{\ast}_{i_2+I_2 \to
f+F}(\mathbf{q^{}_1}+\mathbf{q})\nonumber\end{eqnarray}
\begin{eqnarray} =-i\left[A_{i_1+I_1 \to i_2+I_2}(\mathbf{q})-A^{\ast}_{i_2+I_2
\to i_1+I_1}(-\mathbf{ q})\right] \end{eqnarray}
and entering the abbreviation $S\equiv\exp[i\Phi]$, we find
\begin{eqnarray} \sigma^{tot}(i,I)=2\mathrm{Re}\int d^2b \Big\langle
1-\left\langle\left\langle S \right\rangle \right\rangle\Big\rangle,
 \end{eqnarray}
\begin{eqnarray} \sigma^{tot}_{coh}(i,I)=\int d^2b\Big\langle
1-2\mathrm{Re}\left\langle\left\langle S \right\rangle\right\rangle
+\left\vert\left\langle \left\langle S\right\rangle\right\rangle
\right\vert^2\Big\rangle ,
\end{eqnarray}
\begin{eqnarray} \sigma^{tot}_{incoh}(i,I)=\int d^2b \Big\langle
1-\left\vert\left\langle \left\langle S \right\rangle\right\rangle
\right\vert^2\Big\rangle , \end{eqnarray}
where the double brackets
$\left\langle\left\langle\;\right\rangle\right\rangle$ signify that
averaging is performed over all the configurations of EA and TA in
the i-th and I-th states.

In doing so, the following expressions are valid:
 \begin{eqnarray}\left\langle f \right\rangle=\int d^3r
\,\vert\psi_i(\mathbf{r})\vert^2 f(\mathbf{r})\ ,
\end{eqnarray}
\begin{eqnarray} \left\langle\left\langle F
\right\rangle\right\rangle=\int \prod_{k=1}^{Z}d^3r_k
\,\vert\Psi_I(\{\mathbf{r_k}\})\vert^2 F(\{\mathbf{r_k}\}).
\end{eqnarray}

The relation defining the $\left\langle\left\langle S
\right\rangle\right\rangle$ can be written in an abbreviated form as
\begin{eqnarray} \left\langle\left\langle S \right\rangle\right\rangle
=\exp(i\bar{\Phi}), \end{eqnarray}
where $\bar\Phi(\mathbf{b},\mathbf{s})$ is an effective (`optical')
phase-shift function in the optical model of the full version of the
Glauber theory.

\section{\large Second-order approximation}

In the  so-called optical-model perturbative approximation
\cite{Glauber2}, the `optical' phase-shift function
$\bar\Phi(\mathbf{b})$ can be written as
\begin{eqnarray}\label{expan} \bar\Phi(\mathbf{b},\mathbf{
s})=\sum_{n=1}^{\infty}\frac{i^{n-1}}{n!}\Phi_n,\end{eqnarray}
where
\begin{eqnarray} \Phi_1=\left\langle
\left\langle\Phi\right\rangle\right\rangle,\quad
\Phi_2=\left\langle\left\langle
(\Phi-\Phi_1)^2\right\rangle\right\rangle,\end{eqnarray}
\begin{eqnarray}\Phi_3= \left\langle\left\langle (\Phi-\Phi_1)^3
\right\rangle\right\rangle, \quad \Phi_4= \left\langle\left\langle
(\Phi-\Phi_1)^4 \right\rangle\right\rangle -3\Phi_2^2, \quad \ldots
\nonumber\end{eqnarray}%
\begin{eqnarray}\Phi_n\sim
Z\left(\frac{\alpha}{\beta}\right)^n.\nonumber\end{eqnarray}

The first order for $\bar\Phi(\mathbf{b},\mathbf{s})$ is the double
average of the phase-shift function
$\Phi(\mathbf{b},\mathbf{s},\{\mathbf{s_k}\})$ over all
configurations of EA and TA in the $i$-th and
 $I$-th states. The second-order term of $\bar\Phi(\mathbf{b},\mathbf{s})$ is
purely absorptive and is equal in order of magnitude to the $
Z\alpha^2.$

When the remainder term $R_3(\mathbf{b},\mathbf{ s})$ in the series
(\ref{expan}) is much smaller than unity
\begin{eqnarray} R_3(\mathbf{b},\mathbf{
s})=\sum_{n=3}^{\infty}\frac{i^{n-1}}{n!}\Phi_n\ll 1\
,\end{eqnarray}
it seems natural to neglect them and consider the following
approximation:
\begin{eqnarray}\label{second} \bar\Phi(\mathbf{b},\mathbf{s})\approx
\Phi_1(\mathbf{b},\mathbf{s})+\frac{i}{2}\Phi_2(\mathbf{b},\mathbf{
s}).\end{eqnarray}
The last term  in (\ref{second}) corresponds to the incoherent
scattering.

In order to consider the electron correlations in the TA ground
state, it is useful to define inclusive densities. They can be
defined by integrating over the remaining coordinates
\begin{eqnarray} \rho^{}_{{\scriptscriptstyle Z}-1}(\mathbf{r^{}_1},\ldots
,\mathbf{r^{}_{{\scriptscriptstyle Z}-1}})\equiv\int
d^3r^{}_{\scriptscriptstyle Z}\rho^{}_{\scriptscriptstyle
Z}(\mathbf{r^{}_1},\ldots ,\mathbf{r^{}_{\scriptscriptstyle Z}})
\end{eqnarray}
with
\begin{eqnarray} \rho^{}_{\scriptscriptstyle Z}(\mathbf{r^{}_1},\ldots ,
\mathbf{r^{}_{\scriptscriptstyle Z}})=\vert
\Psi^{}_0(\mathbf{r^{}_1},\ldots ,\mathbf{r^{}_{\scriptscriptstyle
Z}})\vert^2\ .
\end{eqnarray}
Each of these functions is symmetric and normalized
to unity when integrated over all of its coordinates.

In particular, the two-particle and one-particle densities can be
represented as
\begin{eqnarray}\label{rho1} \rho^{}_2(\mathbf{r^{}_1},\mathbf{r^{}_2})=\int
d^3r^{}_3\rho^{}_3(\mathbf{r^{}_1},
\mathbf{r^{}_2},\mathbf{r^{}_3}), \quad
\rho^{}_1(\mathbf{r^{}_1})=\int
d^3r^{}_2\rho^{}_2(\mathbf{r^{}_1},\mathbf{r^{}_2})\ .
\end{eqnarray}
The two-particle density $\rho^{}_2(\mathbf{r^{}_1},\mathbf{
r^{}_2})$ describes the probability of findings any two of the
properly antisymmetrized electrons at positions $\mathbf{r^{}_1}$
and $\mathbf{r^{}_2}$.

Taking a Fourier transform, we obtain the one-particle $\tilde
F_1(\mathbf{q})$ and two-particle $\tilde
F_2(\mathbf{q}_1,\mathbf{q}_2)$ TA form factors, which are just the
expectation values of special one-particle and two-particle
operators
\begin{eqnarray}
  \label{eq:648}
\tilde F^{}_1(\mathbf{q})&\equiv&\int d^3r^{}_1
e^{i\mathbf{q}\mathbf{r^{}_1}} \rho^{}_1(\mathbf{r^{}_1})\,,
\end{eqnarray}
\begin{eqnarray}
  \label{eq:649}
\tilde F^{}_2(\mathbf{q^{}_1},\mathbf{q^{}_2})&\equiv&\int d^3r^{}_1
d^3r^{}_2
e^{i\mathbf{q^{}_1}\mathbf{r^{}_1}-i\mathbf{q^{}_2}\mathbf{r^{}_2}}
\rho^{}_2(\mathbf{r^{}_1},\mathbf{r^{}_2})\, .
\end{eqnarray}

All the many-particle densities can be expressed in terms of
one-particle static and transition densities. Using  canonical
anticommutation relations one can immediately establish the
following relations for the correlation term
$W(\mathbf{q_1},\mathbf{q_2})$:
\begin{eqnarray}
  \label{eq:647}
W(\mathbf{q_1},\mathbf{q_2})&=& \tilde
F_1(\mathbf{q_1}-\mathbf{q_2})-\tilde F_1(\mathbf{q_1})\tilde
F_1(\mathbf{q_2})\nn &&+(Z-1)
[\tilde F_2(\mathbf{q_1},\mathbf{q_2})-\tilde F_1(\mathbf{q_1})\tilde F_1(\mathbf{q_2})]\,,\\
  \label{eq:651} &&~~~~~~~~~~~~~~~~W(\mathbf{q},\mathbf{q})=\tilde
F_{incoh}(\mathbf{q})\, .
\end{eqnarray}

Finally, putting $\mathbf{b_{\pm}} =\mathbf{b} \pm \mathbf{s}/2$, we
express the quantities
 $\Phi_1(\mathbf{b},\mathbf{s})$ and $\Phi_2(\mathbf{b},\mathbf{s})$ as
\begin{eqnarray}
  \label{eq:645}
\Phi_1=\frac{2Z\alpha}{\beta} \int \frac{d^2q}{q^{\,2}} \left(
e^{i\mathbf{q}\mathbf{b}_+} -e^{i\mathbf{q}\mathbf{b_-}}\right)
[1-\tilde F_1(\mathbf{q})]\,,\\
  \label{eq:646}
\Phi_2=\frac{4Z\alpha^2}{\beta^2} \int \frac{d^2q^{}_1}{q_1^{\,2}}
\frac{d^2q^{}_2}{ q_2^{\,2}} \left(
e^{i\mathbf{q^{}_1}\mathbf{b^{}_+}}
-e^{i\mathbf{q^{}_1}\mathbf{b^{}_-}}\right)\!\!\left(e^{-i\mathbf{q^{}_2}\mathbf{b^{}_+}}
-e^{-i\mathbf{q^{}_2}\mathbf{b^{}_-}}\right)
\end{eqnarray}
$$~~~~~~~~~~~~~~~~~~~~\times W(\mathbf{q^{}_1},\mathbf{q^{}_2})\ .$$
Let us notice that these expressions are in agreement with the
preliminary results of \cite{Proc}.

Making use of the relations
\begin{eqnarray} \sigma_{coh}^{tot}(i,I)
=\left\langle\sigma_{coh}^{tot}(\mathbf{ s})\right\rangle, \quad
\sigma_{incoh}^{tot}(i,I)
=\left\langle\sigma_{incoh}^{tot}(\mathbf{s})\right\rangle,
\end{eqnarray}
\begin{eqnarray} \sigma^{tot}(i,I) =\left\langle\sigma^{tot}(\mathbf{
s})\right\rangle\ ,\nonumber \end{eqnarray}
we can find the following expressions for all `dipole total cross
sections' $\sigma^{tot}_{coh(incoh)}(\mathbf{s})$, depending only on
the properties of the target material:
\begin{eqnarray} \sigma^{tot}(\mathbf{s})=2 \int d^2b
\left(1-\cos\Phi_1^{}e^{-\Phi^{}_2/2}\right),\end{eqnarray}
\begin{eqnarray} \sigma^{tot}_{coh}(\mathbf{s})=\int  d^2b
\left(1-2\cos\Phi_1^{}e^{-\Phi^{}_2/2}+e^{-\Phi^{}_2}\right),\end{eqnarray}
\begin{eqnarray} \sigma_{incoh}^{tot}(\mathbf{s})=\int  d^2b
\left(1-e^{-\Phi^{}_2}\right).\end{eqnarray}

To establish a connection between the results obtained in this  work
and in \cite{Tarasov,Heim}, we rewrite the total cross sections of
the EA--TA interactions
\begin{equation}
  \label{eq:641}
  \sigma^{tot}=\sigma^{tot}_{coh}+\sigma^{tot}_{incoh}
\end{equation}
in terms of the interaction operators
$\Gamma_{coh(incoh)}(\mathbf{b},\mathbf{s})$
\begin{equation}
  \label{eq:642}
  \sigma^{tot}_{coh(incoh)}=\int d^3r |\Psi_{i(I)}(\mathbf{r})|^2
  d^2b\,
  \Gamma_{coh(incoh)}(\mathbf{b},\mathbf{s}) \ ,
\end{equation}
where $\sigma^{tot}_{coh(incoh)}$ are the total cross sections of
the EA--TA interaction with or without excitation of the target
atom. In (\ref{eq:642}), we applied the abbreviation
$\int\prod_{k=1}^{Z}d^3r_k
\,\vert\Psi_I(\{\mathbf{r_k}\})\vert^2\equiv\int d^3r
\,\vert\Psi_I(\mathbf{r})\vert^2 $ and operators who reads
\begin{equation}
  \label{eq:643}
  \Gamma_{coh}(\mathbf{b},\mathbf{s})=1-
  2\cos{[\Phi_1(\mathbf{b},\mathbf{s})]}\exp{[-\Phi_2(\mathbf{b},\mathbf{s})/2]}
 + \exp{[-\Phi_2(\mathbf{b},\mathbf{s})]}\,,
\end{equation}
\begin{equation}
  \label{eq:644}
  \Gamma_{incoh}(\mathbf{b},\mathbf{s})=1 - \exp{[-\Phi_2(\mathbf{b},\mathbf{
  s})]}\,.
\end{equation}
In the above equations, the functions
$\Phi_1(\mathbf{b},\mathbf{s})$ and $\Phi_2(\mathbf{b},\mathbf{s})$
are given by  (\ref{eq:645}) and (\ref{eq:646}). The phase-shift
function $\Phi_2$ accounts for the TA excitations both in the
intermediate and final states. At $\Phi_2=0$, the expressions
(\ref{eq:641})--(\ref{eq:644}) can be reduced to the corresponding
relations of refs. \cite{Tarasov,Heim}. In particular,
$\sigma^{tot}_{incoh}=0$ in this limit.

\section{\large `Potential' approximation of the second-order optical model}

The eikonal approximation for EA--TA scattering neglecting effects
of the intermediate excitations of TA (`potential' approximation)
can be represented as follows:
\begin{eqnarray} \left[\sigma_{incoh}^{tot}(i,I)\right]_{pot}\approx 0, \quad
\left[\sigma^{tot}(i,I)\right]_{pot}\approx\left[\sigma_{coh}^{tot}(i,I)\right]_{pot}.
\end{eqnarray}

Let us define the absolute accuracy of this approximation as
\begin{eqnarray} \Delta\sigma_{coh}^{tot}(i,I)&\equiv&
\sigma_{coh}^{tot}(i,I)-\left[\sigma_{coh}^{tot}(i,I)\right]_{pot}
=\left\langle\Delta\sigma_{coh}^{tot}(\mathbf{s})\right\rangle\nonumber
 \end{eqnarray}
with
\begin{eqnarray}
\sigma_{coh}^{tot}= \sigma^{to
t}-\sigma_{incoh}^{tot}
=\sum_{f}\sum_{F}\sigma_{i+I \to f+F}-\sum_{f}\sum_{F\ne
I}\sigma_{i+I \to f+F},\nonumber\end{eqnarray}
\begin{eqnarray}\left[\sigma_{coh}^{tot}(i,I)\right]_{pot}\approx
\left[\sigma^{tot}(i,I)\right]_{pot} =\sum_{f}\sigma_{i+I \to f+I}.
\end{eqnarray}
Within the second-order perturbation theory, one gets the following
expression for this quantity:
\begin{eqnarray}
\Delta\sigma_{coh}^{tot}(\mathbf{s})&=&\sigma_{coh}^{tot}(\mathbf{
s})-\left[\sigma_{coh}^{tot}(\mathbf{s})\right]_{pot}\nonumber\\
&=&\int d^2b
\left[e^{-\Phi^{}_2}-1+2(1-\cos\Phi_1^{})e^{-\Phi^{}_2/2}\right].
 \end{eqnarray}
Here, the phase-shift functions $\Phi_1^{}$ and $\Phi_2^{}$ are
defined by (\ref{eq:645}) and (\ref{eq:646}).

To estimate the other corrections, we will use the evaluation
formulae given by:
\begin{eqnarray} \int \Phi_1^2(\mathbf{b},\mathbf{s})\;d^2b\sim
(Z\alpha)^2\,s^2\,L\ , \quad \int\Phi_1^{2k}(\mathbf{b},\mathbf{
s})\;d^2b\sim (Z\alpha)^{2k} s^2 \ ; \end{eqnarray}
\begin{eqnarray} \label{Za2} \int \Phi_2(\mathbf{b},\mathbf{s})\;d^2b\sim
(Z\alpha^2)\,s^2\,L \ , \quad \int \Phi^2_2(\mathbf{b},\mathbf{
s})\;d^2b\sim (Z\alpha^2)^2\, \frac{s^4}{R^2_+}\,L^2 \ ;
\end{eqnarray}
\begin{eqnarray} \label{Z3a4}\int
\Phi^2_1(\mathbf{b},\mathbf{s})\;\Phi_2(\mathbf{ b},\mathbf{s})\;
d^2b\sim (Z^3\alpha^4)\,\frac{s^4}{R^2_+}\,L^2 \ , \nonumber
\end{eqnarray}
\begin{eqnarray}\int \Phi^{2k}_1(\mathbf{b},\mathbf{s})\;
\Phi_2(\mathbf{b},\mathbf{ s})\; d^2b\sim
(Z\alpha)^{2k}\;(Z\alpha^2)\;\frac{s^4}{R^2_+}\,L \end{eqnarray}
with
\begin{eqnarray} L=\ln\frac{R^2_+}{s^2} \ , \quad \mathbf{ R_+}=\mathbf{
R}+\frac{\mathbf{r}}{2} \ , \quad k\geq 1 \ .\end{eqnarray}

Using the  definition
$$\bar L=\ln\frac{R^2_+}{\left\langle s^2\right\rangle}$$
and the  evaluation formulae (\ref{Za2}), we find the following
relation between the total cross sections of the incoherent
scattering in the Glauber and Born approximations:
\begin{eqnarray} \sigma^{tot}_{incoh}=\left[\sigma^{tot}_{incoh}
\right]_{Born}\left[1+ O\left(Z\alpha^2\,\frac{\left\langle
s^2\right\rangle}{R^2_+}\,\bar L\right)\right], \end{eqnarray}
where
\begin{eqnarray} \left[\sigma^{tot}_{incoh} \right]_{Born}=\left\langle\int
d^2b\; \Phi_2(\mathbf{b},\mathbf{s})\right\rangle. \end{eqnarray}
The difference between the first-order and second-order total cross
sections of the incoherent scattering normalized to the first-order
cross section reads:
\begin{eqnarray} \label{Born}
\frac{\sigma^{tot}_{incoh}}{\left[\sigma^{tot}_{incoh}
\right]_{Born}}-1\equiv\frac{\Delta\sigma^{tot}_{incoh}}{\left[\sigma^{tot}_{incoh}
\right]_{Born}} = O\left(Z\alpha^2\,\frac{\left\langle
s^2\right\rangle}{R^2_+}\,\bar L\right)\ . \end{eqnarray}
It follows from (\ref{Born}) that the incoherent interactions can be
described by the Born approximation with a relative accuracy of the
order of $Z\alpha^2$.  In terms of the average radii of the
interacting objects, they can be presented as
\begin{equation}
 Z\alpha^2\; \frac{\langle r^2\rangle_{EA}}{\langle
r^2\rangle_{TA}} \ln\left(\frac{\langle r^2\rangle_{TA}}{\langle
r^2\rangle_{EA}}\right) \,.
\end{equation}
The obtained result shows that the Born approximation used in
\cite{Basel} to describe  the incoherent sector of the
$A_{2\pi}$--TA interactions is sufficiently accurate in the context
of the DIRAC experiment.

From (\ref{Z3a4}),  it follows that the relative correction to the
DMA--TA interaction cross section $\sigma^{tot}_{coh}(i,I)$ provided
by the intermediate incoherent effects is of the order of
\begin{equation}
\label{positronium}
 Z^3\alpha^4\; \frac{\langle r^2\rangle_{EA}}{\langle
r^2\rangle_{TA}} \ln\left(\frac{\langle r^2\rangle_{TA}}{\langle
r^2\rangle_{EA}}\right)\ll 1
\end{equation}
and can be safely neglected. This agrees with the conclusion of
\cite{Basel} done on the basis of more rough estimations. The same
is true for all partial coherent cross sections. This result
indicates that the theory of refs. \cite{Tarasov,Heim} provides
quite an accurate description for the coherent sector of the DMA--TA
interactions.

Let us notice that the mentioned discrepancy between the  results
obtained for the $A_{2\pi}$--TA \cite{Basel} and $A_{2e}$--TA
\cite{Pak} interactions is a result of expression
(\ref{positronium}), since $\langle r^{2}\rangle_{2e}\gg \langle
r^{2}\rangle_{2\pi}$.

For the elastic scattering
\begin{eqnarray} \sigma^{el}_{i+I \to i+I}(i,I)= \int d^2q \vert A_{i+I \to
i+I}(\mathbf{q})\vert^2 \ ,\end{eqnarray}
we also obtain a relation to its `potential' approximation:
\begin{eqnarray} \sigma^{el}_{i+I \to i+I}=\left[\sigma^{el}_{i+I \to
i+I}\right]_{pot}\left(1+ \frac{1}{Z}\,\frac{\left\langle
s^2\right\rangle}{R^2_+}\,\bar L\right). \end{eqnarray}
The relative accuracy of this approximation can be estimated as
\begin{eqnarray} \frac{\sigma^{el}-
\left[\sigma^{el}\right]_{pot}}{\left[\sigma^{el}\right]_{pot}}
\equiv\frac{\Delta\sigma^{el}_{i+I \to i+I} }{\left[\sigma^{el}_{i+I
\to i+I}\right]_{pot}}=\frac{1}{Z}\,\frac{\left\langle
s^2\right\rangle}{R^2_+}\,\bar L \ .\end{eqnarray}

For the purposes of the DIRAC experiment, the results of the
performed analysis can be summarized as follows: (i) for the
description of the coherent DMA--TA interactions, it is enough to
use a simplified version of the Glauber theory \cite{Tarasov,Heim},
which neglects the effects of the intermediate TA excitations; (ii)
for the description of the incoherent DMA--TA interactions, it is
enough to use the Born approximation. This analysis substantiates
the use of the `potential approximation' for the second-order
optical model in the DIRAC experimental data processing
\cite{Dirac2,Dirac3}; and, it has recently shown, it allows one,
among other things, to achieve the declared accuracy of 10\% in
determining the $A_{2\pi}$ lifetime \cite{Dirac3}.

\section{\large Conclusion}
In this work, a complete  version of the Glauber theory is
formulated for the EA--TA scattering  accounting all possible
excitations of EA and TA in intermediate and/or finale states. Its
second-order optical model is analyzed. In the framework of this
model, the accuracy of the `potential' approximation is evaluated.

The work gives a natural generalization of the conventional Glauber
theory for high-energy  scattering of relativistic hydrogenlike
elementary atoms\footnote{One can enumerate $A^{}_{2\pi}$,
$A^{}_{\pi \scriptscriptstyle K}$, $A^{}_{\scriptscriptstyle KK}$;
$A^{}_{e\pi}$, $A^{}_{\mu\pi}$, $A^{}_{e K}$, $A^{}_{\mu K}$;
$A^{}_{2e}$, $A^{}_{e\mu}$, $A^{}_{2\mu}$; $A_{p\pi}$, $A_{pK}$,
$A_{p\mu}$, $A_{pe}$ here.} by target atoms\footnote{Applied to the
experiment DIRAC, we examined primarily Be, Al, Ti, Ni, Mo, Sn, Ta,
Pt, Au, Pb, etc.}. We would like to note that while the theory
developed in this work is motivated by a specific experiment
(DIRAC), it is also of more general interest for high energy physics
and atomic physics.

\section*{\large Acknowledgments}

I would like to express my gratitude to Sergey Gevorkyan and Marina
Aristarkhova for their thoroughly proofreading of the manuscript and
useful comments.

\end{document}